# Numerical investigation into the injection-locking phenomena of gain switched lasers for optical frequency comb generation

S. P. O'Duill, P. M. Anandarajah, R. Zhou and L. P. Barry

*The RINCE Institute, Dublin City University, Dublin 9, Ireland*

Corresponding author: *sean.oduill@dcu.ie*

We present detailed numerical simulations of the laser dynamics that describe optical frequency comb formation by injection-locking a gain-switched laser. The typical rate equations for semiconductor lasers including stochastic carrier recombination and spontaneous emission suffice to show the injection-locking behavior of gain switched lasers, and we show how the optical frequency comb evolves starting from the free-running state, right through the final injection- locked state. Unlike the locking of continuous wave lasers, we show that the locking range for gain switched lasers is considerably greater because injection locking can be achieved by injecting at frequencies close to one of the comb lines. The quality of the comb lines are formally assessed by calculating the FM-noise spectral density and we show that under injection-locking conditions the FM-noise spectral density of the comb lines tend to that of the maser laser.

Optical frequency combs (OFC) are necessary to perform many signal processing tasks [1] and for high capacity optical communication networks [3], [3]. One method to create OFCs is by gain-switching a semiconductor laser [4]-[10]. The gain switching of the laser using a high power sine wave signal that drives the laser gain above and below the lasing threshold is a convenient method to create a train of short optical pulses which in the frequency domain is an optical frequency comb [4]. The frequency spacing between each comb line, also known as the free-spectral range (FSR), is determined by the frequency of the sinusoidal modulation. OFCs from gain switched lasers are extremely versatile, offering tunability of the central wavelength (if the laser allows for such functionality) [4], [7] [8] and the FSR [6]. In conjunction with the injection-locking technique, gain switched lasers have been shown to have an impressive information carrying capacity exceeding 2 Tbit/s with a net spectral efficiency reaching 6.7 bit/s/Hz [11]. In addition to being employed in high capacity networks, the combination of wavelength and FSR tunability makes gain-switched lasers particularly ideal sources for use within future flexible grid optical networks [12].

Injection-locking of gain-switched lasers, using a continuous wave (CW) master laser dramatically improves the quality of the comb [4] [7],[8] i.e. a well-defined optical frequency comb appears in the output spectrum as well as controlling the phase noise properties (or the spectral line-broadening) of each comb line. A schematic of the process is shown in Fig. 1.

The slave laser is biased above threshold and a sinusoidal modulating current is coupled with the bias current to gain-switch the slave laser. A master laser is operated in CW mode and injected into the slave laser, the combination of the gain switching and external injection creates a comb with the FSR determined by the frequency of the sinusoidal modulation and phase noise determined by that of the master laser when the injection locking condition is satisfied. While extensive experimental work has been carried out to understand the operating principles and development of the gain-switched lasers [4]-[10], to date little numerical investigation has been performed to investigate the injection locking mechanism of gain-switched lasers for OFC generation apart from showing that laser rate equations can simulate a comb output with similar spectral shape to experimental results [9], though no results pertaining to the spectral quality of the comb lines nor to the unlocked phenomena were presented.

In this letter, we use a stochastic semiconductor laser rate equation model to investigate the injection-locking properties of gain-switched lasers. We detail the injection locking behavior by examining the output spectra as the power of the injected master signal is increased. We notice that the comb is redshifted relative to the free running comb without external injection, and as the injection power is increased the comb continues to redshift until one of the comb lines directly coincides and locks to the master laser. The FSR of the comb is determined by the modulating signal, with comb line frequencies satisfying the condition that $f_n = f_{mas} + n\Delta f_{mod}$, where $f_n$ is the frequency of the $n^{th}$ comb line, $f_{mas}$ is the lasing frequency of the master laser, $\Delta f_{mod}$ is the RF modulating frequency and $n$ is an integer. We also show that injection-locking of the comb is possible by injecting close to any of the comb lines in the free running comb offering injection-locking over a much wider frequency range as compared with injection-locking of CW lasers [13]-[15], such a feature is unique to these gain-switched lasers. The spectral quality of the comb lines is determined by phase noise exhibited by the comb lines. Hence, we calculate the FM-noise spectral density of the lines to show that FM noise of the comb lines reduces to that of the master laser as the power of the injected master laser is increased. In our recent work on the modelling of gain-switched lasers [9], we presented results showing that the laser rate equations can simulate OFCs observed from experiments. However, the numerical technique employing three separate, coupled rate equations for the laser carrier density, photon density and phase was unstable and the technique could only generate numerical results in the absence of external injection or when an injection-locked comb was created. A different approach is sought to undertake the task set out in this letter and we resort to the optical field representation of the laser rate equation [13], with $E$ being the complex optical field of the laser such that $|E|^2$ represents the photon density. In this case the photon density never goes negative and causes no issues in the numerical calculations irrespective of the system parameters. The two laser rate equations for the carrier density $N$ and optical field $E$, complete with stochastic terms needed to describe the system are:

$$\frac{dN}{dt} = \frac{I(t)}{eV} - R(N) - a(N - N_0)|E|^2 + F_N \quad (1)$$

and

$$\frac{dE}{dt} = \frac{1}{2}\left[(1 - j\alpha_H)a(N - N_0) - \frac{1}{\tau_p}\right]E + j\frac{\alpha_H}{2\tau_p}E + k_c E_{inj} + F_E \quad (2)$$

Where all of the symbols have their usual meanings and are defined in Table I. The first term on the right hand side of (1) is given by the electric current flowing into the laser that includes the gain-switching current $I(t) = I_{bias} + I_{mod}\sin(2\pi\Delta f_{mod} t)$ where $I_{bias}$ = 75 mA, $I_{mod}$ = 75 mA and $\Delta f_{mod}$ = 12.5 GHz throughout; carrier recombination $R(N) = AN + BN^2 + CN^3$ for nonradiative, bimolecular and Auger recombination respectively; the third term represents stimulated emission; and $F_N$ denotes stochastic carrier recombination (which will be defined later). The first term on the right hand side of (2) describes the complex gain of the laser field including the losses. The second term $j(\alpha_H/2\tau_p)E$ is a constant frequency shift of the laser field, the inclusion of this term sets the central frequency of $E$ to equal zero when the laser is operated in CW mode above threshold. The inclusion of this term is crucial to exactly defining the frequencies of the slave laser and the maser laser, so that the maser–slave detuning can be simply defined by $\Delta f_{det}$ in the term for $k_c E_{inj}$ with $E_{inj} = E'_{inj}\exp(j2\pi\Delta f_{det} t)$, where $E'_{inj}$ is the field of the master laser. The final term is the random addition of spontaneous emission due to bimolecular recombination into the lasing field. $F_E$ denotes spontaneous emission into the lasing field.

The stochastic terms are given by:

$$F_N = \sqrt{2R(N)B_{sim}}\, e_N(t) \quad (3)$$

$$F_E = \sqrt{\beta BN^2 B_{sim}}\,(e_{EI}(t) + je_{EQ}(t)) \quad (4)$$

Where $B_{sim}$ is the simulation bandwidth; $e_N$, $e_{EI}$ and $e_{EQ}$ are the stochastic Gaussian random variable waveform terms with zero mean and unity variance. $e_{EI}$ and $e_{EQ}$ describe the phase and quadrature components of the spontaneous emission into the laser field. The value of $\beta$ controls the amount of spontaneous emission into the lasing field and can therefore can control the phase noise of the output field. In order to show linewidth reduction of the slave laser, we generate the slave laser with a value of $\beta = 5\times10^{-4}$ and for the master laser with a value of $\beta = 5\times10^{-5}$. The optical power is a scaled version of the photon density $P = K|E|^2 = K'v_g A_{eff} h\upsilon |E|^2$, with $\upsilon_g$ being the group velocity, $A_{eff}$ being the effective transverse area of the laser field, $h\upsilon$ is the photon energy and $K' = 0.3$ is a general scaling factor that can be adjusted depending on the type of single mode laser used, e.g. distributed feedback lasers and discrete mode lasers [4]. The optical spectrum is given by $|\tilde{E}(\omega)|^2$ where $\tilde{E}(\omega) = F(\sqrt{K}E)$ and $F$ is the Fourier transform operator.

The field for the master laser is generated using (1) and (2) with the same values given in Table I by only applying the DC bias current of 75 mA, setting the injection term to zero and setting $\beta = 5\times10^{-5}$. To avoid frequency placement uncertainties, it is best to display all of the spectral results with the zero frequency to be the central frequency of the master laser. The frequency detuning between the master and the slave (both under CW conditions) is set to $\Delta f_{det} = f_{mas} - f_{slave}$ = -3 GHz; unless otherwise stated all of the spectral results obtained via the simulation are displayed with the frequency axis translated by +3 GHz. In order to obtain an insight into the gain switching process we begin by presenting results for the case when there is no light injected from the master into the gain-switched slave laser using a 12.5 GHz sine wave. The results are shown in Fig. 2. The spectrum of the free running comb is shown in Fig. 2(a), an OFC is observed with comb lines separated by 12.5 GHz apart and with the central comb line at +3 GHz, which corresponds to the frequency of the slave laser in CW mode. When we inject 8 μW from the master laser into gain switched laser the gain switched comb is injection locked and the result is shown in Fig. 2(b), the entire comb has now has been shifted by -3 GHz such that one comb line coincides with the CW frequency of the master laser and all of the other lines in the spectrum correspond to an OFC with an FSR of 12.5 GHz. The conditions for creating an OFC under external injection conditions are satisfied in (1) and (2) only when $N$ and $E$ are periodic. Therefore the condition that supports $N$ being periodic is satisfied when $|E|^2$ is periodic, thus this in turn implies that a term proportional to $E_{inj}$ must constitute part of the comb output. The locking of the comb is accompanied by a dramatic reduction in the line broadening of each line, because the linewidth of the master laser is set to be an order of magnitude smaller than that of the slave laser, the same was observed for the locking of CW lasers [14]. The linewidth narrowing is shown in Fig. 2(c) though later we formally evaluate the line broadening by calculating the FM-noise spectral density which gives clearer insight into the phase noise reduction of the slave.

Next we show what happens at intermediary power levels to show how the locking process evolves from the free-running to the injection-locked case. The simulations were performed by gradually increasing the injected power from the master laser and recording the output spectra. A selection of the spectra at increasing injection power levels are shown in Fig. 3. A zoom of the optical spectra about the zero frequency suffices to display the locking mechanism. Fig. 3(b)–(e) show the spectra when an insufficient amount of power is injected into the slave laser to cause locking, these spectra show three distinct features (all labeled in Fig. 3(b)): there are two combs present in the output spectrum, one is due to gain-switching of the slave laser, and this comb has been slightly redshifted from the free running case due to the interaction of the comb with the injected master. A comb due to the injected master field appears in the spectrum and is caused by the injected master field being gain-modulated due to gain-switching. Components of this comb appear at frequencies $f_{mas} \pm n\Delta f_{mod}$ (which happen to be the frequencies of the injection-locked comb). Finally, four-wave mixing (FWM) components due to the beating between a line from the comb from the slave with the closest frequency from the modulated fields of the injected master. The presence of the FWM terms indicate the absence of locking though there is a beating between the comb and the injected fields. The redshifting of the comb from the slave can be explained by the injection locking of CW lasers, that the coherent addition of the master field with the slave field requires less average population inversion ($N$) to sustain lasing [14] and hence reduces the average emission frequency to that of the master, allowing phase locking to occur. Our simulations show that $N$ reduces from $1.4165 \times 10^{24}$ m$^{-3}$ to $1.399 \times 10^{24}$ m$^{-3}$ between the cases of the free running and injection locked combs.

As the power of the injected master is increased, the comb from the slave gets closer to the master comb and then the slave comb finally phase locks to the master field, such that tendencies of the slave laser to exhibit large phase fluctuations are suppressed by the master. At this point we now have a comb from the slave laser whose FSR is controlled by the applied sinusoidal modulation, and the phase noise of the individual comb lines is determined by that of the injected master laser.

We next explore the master-slave detuning possibilities for locking of gain-switched lasers. The requirement that the master constitutes one of the comb lines suggests that if one was able to tune the optical frequency of the master laser to be close to one of the other comb lines then the comb should be injection-locked to the master. To show this, we increase the detuning between the CW master and CW slave laser to be 15.5 GHz instead of 3 GHz (one whole FSR larger than the case outlined above) such that the master is detuned by -3 GHz from the first red-shifted comb line relative to the central comb line of the free running comb. The power of the master set to 7 µW. The comb spectrum of the injection-locked comb is shown in Fig. 4, which shows that the comb is locked to the master and all comb lines satisfy the criteria that they exist at frequencies of $f_{mas} \pm n\Delta f_{mod}$. This shows that there exists multiple locking ranges for stable injection locking of gain-switched lasers within the spectral regions of the most powerful comb lines and that the locking capabilities are remarkably different (and superior) than for the locking of CW lasers [14],[15] where only a narrow locking range exists.

The phase noise of the comb lines is quantified by the FM-noise spectral density [8]. A comb line is filtered out using a Gaussian-shaped optical bandpass filter with 2 GHz bandwidth which is sufficient to reject the adjacent comb lines. This creates a CW field from which we extract the temporal phase noise, $\Delta\phi$, and convert to frequency through differentiation to give a frequency deviation waveform $\Delta f_n$. The FM noise spectral density is given by the Fourier transform of the autocorrelation of $\Delta f_n$.

$$S_{\Delta f_n} = F\left[\int_{-\infty}^{\infty} \Delta f_n(t) \Delta f_n(t+\tau) d\tau\right] \tag{5}$$

The FM-noise spectral density plots are calculated for the free-running comb, master laser and for injection-locked combs when the power of the injected master is 6, 8 and 20 µW. The upper and lower bounds of the FM spectral density are the values from the gain-switched free-running slave and from the injected maser respectively. For the free running slave the gain-switching causes the line broadening of the comb lines to be larger than that of the free running CW slave due to the timing jitter as the pulses are building up from ASE, and due to the fact that the comb lines are weaker than the CW field, thus the ASE creates a larger phase noise [16]. As the power of the injected master is increased, the FM-noise spectral density of the comb lines decreases and tends to that of the master at low frequencies which is further evidence of injection locking because the master laser suppresses frequency fluctuations of the slave laser. The increasing FM-noise spectral density at frequencies > 700 MHz show that the slave is not fully locked to the maser at all frequencies, and displays a tendency to display phase noise given by the ASE. This is in accordance with experimental findings for the FM-noise characteristics of comb lines from injection-locked gain switched lasers in [8].

In conclusion, we have presented a numerical model that is capable of demonstrating the injection-locking properties of gain-switched lasers for OFC generation. This will allow us to optimize the relevant laser design parameters in order to tailor the OFC for any given application.

**Acknowledgements:**

We acknowledge support from Science Foundation Ireland under PI grant Nos. (09/IN 1/I2653), (14/TIDA/2405), (10/CE/I1853) and (12/RC/2276); Enterprise Ireland COMBPIC project; and the HEA (Ireland) PRTLI 4 INSPIRE Programs.

**References**:

[1] P. J. Delfyett, et al., "Optical Frequency Combs From Semiconductor Lasers and Applications in Ultrawideband Signal Processing and Communications," IEEE J. of Lightw. Technol., 24, No. 7, pp 2701 – 2718, 2006.

[2] A. D. Ellis, et al., "Spectral density enhancement using coherent WDM," IEEE Photon. Technol. Lett. 17, No. 2, pp. 504-506, 2005.

[3] D. Hillerkuss et al., "26 Tbit s$^{-1}$ line-rate super-channel transmission utilizing all-optical fast Fourier transform processing," Nat. Photon. 5, pp 364 – 371, 2011.

[4] P. Anandarajah et al., "Generation of Coherent Multicarrier Signals by Gain Switching of Discrete Mode Lasers," IEEE Photon. Journ. 3, No. 1, pp 112 – 122, 2011.

[5] S. Chen, et al.," Spectral dynamics of picosecond gain-switched pulses from nitride-based vertical-cavity surface-emitting lasers," Sci. Reports, 4, Article No. 4735, 2014.

[6] P. Anandarajah et al., "Flexible optical comb source for super channel systems," Proc of OFC 2013, OTh3I8, Los Angeles, 2013.

[7] R. Zhou et al., "40nm Wavelength Tunable Gain-Switched Optical Comb Source," OSA Optics Expr. 19, No. 26, pp. B415-B420, 2011.

[8] R. Zhou et al., "Phase noise analysis of injected gain switched comb source for coherent communications," Opt. Expr.. 22, No. 7, pp.8120-8125, 2014.

[9] S. Ó Dúill, et al, "Numerical Investigation into the Dynamics of Externally-Injected, Gain-Switched Lasers for Optical Comb Generation," Proc. of Euro. Conf. on Opt. Commun. 2014, P.8.1, 2014 Cannes.

[10] A. R. Criado Serrano, et al.," VCSEL-based optical frequency combs: toward efficient single-device comb generation," IEEE Photon. Tech. Lett. 25, No. 20 pp. 1981 – 1984, 2013.

[11] J. Pfeifle et al., "Flexible terabit/s Nyquist-WDM super-channels using a gain-switched comb source," Opt. Expr. 23, No. 2, pp.724-738, 2015.

[12] I. Tomkos et al., "A Tutorial on the Flexible Optical Networking Paradigm: State of the Art, Trends, and Research Challenges," Proc. of the IEEE, 102, No. 9, pp 1317 – 1337, 2014.

[13] N. Schunk and K. Petermann, 'Noise analysis of injection-locked semiconductor injection lasers,' J. Quantum Electron., 22, pp. 642-650, May 1986.

[14] R. Lang, "Injection locking properties of a semiconductor laser," J. of Quantum Electron.,36, No. 6, pp 976 – 983, 1982.

[15] V. Annovazzi-Lodi et al., "Dynamic Behavior and Locking of a semiconductor laser subjected to external injection," J. of Quantum Electron., 34, No. 12, pp 2350-2357, 1998.

[16] P. Anandarajah et al., "Enhanced Optical Comb Generation by Gain Switching a Single Mode Semiconductor Laser Close to its Relaxation Oscillation Frequency," Under review in IEEE J. of Sel. Topics in Quantum Electron.

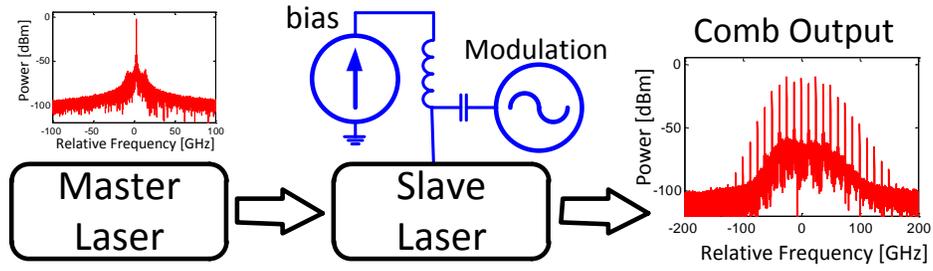

Fig. 1 Schematic of an injection-locked gain switched laser. The slave laser is biased above threshold and modulated using a strong sinusoidal modulation. Light from a CW master laser is injected into the slave laser to that locks the comb output. An example of the comb spectrum is shown at the output of the slave laser.

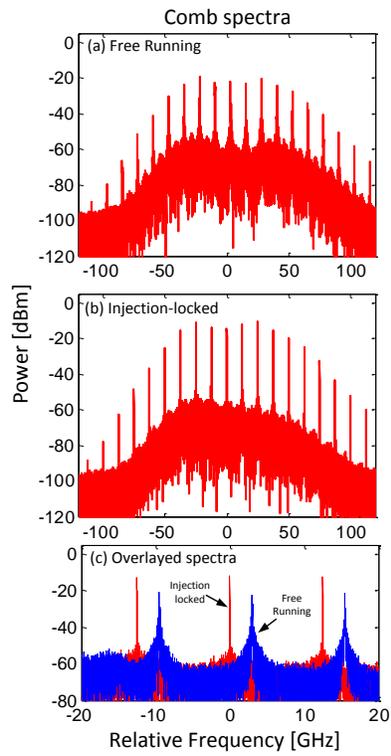

Fig. 2 Calculated comb spectra for the case (a) gain switched free running slave and (b) injection-locked gain-switched slave. The comb lines of the injected locked comb exhibits a smaller linebroadening commensurate with the linebroadening of the master laser. (c) Overlayed spectra of the free-running and injection locked comb to highlight the spectral narrowing of the comb lines.

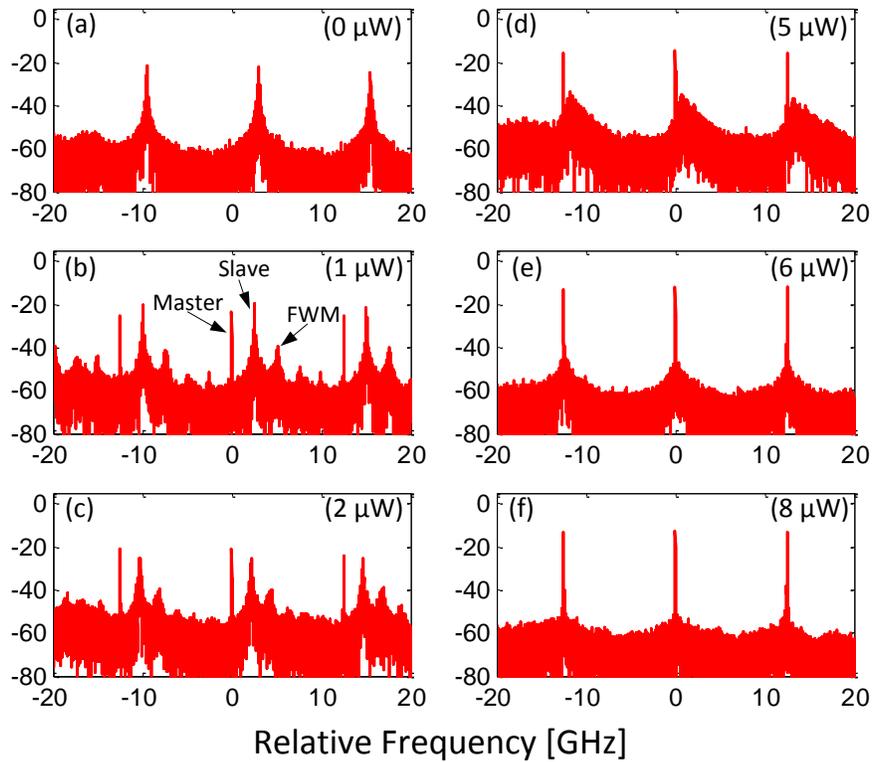

Fig. 3 Plots of the output spectra with increasing power of the injected master. For the case (b) – (d) the comb is not locked to the master laser. In each of these cases, the gain-switched slave laser creates a comb marked 'slave' in (b), the gain switching also causes the injected master to form a comb marked 'master' in (b). As the power is increased (c)-(e), the comb from the slave laser is drawn to the lines of the modulated master laser field and eventually all of the comb lines are locked to the master laser in (e). Increasing the injection power further beyond the case in (e) shows a further narrowing of the comb lines in (f).

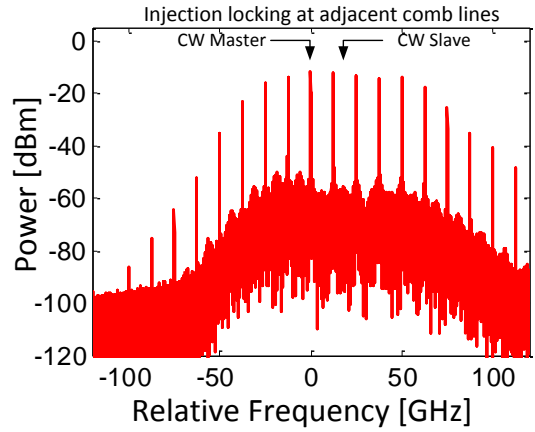

Fig. 4. Simulated comb spectrum when the master-slave detuning is set to 15.5 GHz, which one whole FSR greater than the case shown in Fig. 2. The spectral placement of the comb lines along with the narrowing of each comb line shows that the comb is injection-locked to the master laser. The spectral placement of the CW master and CW slave are indicated

.

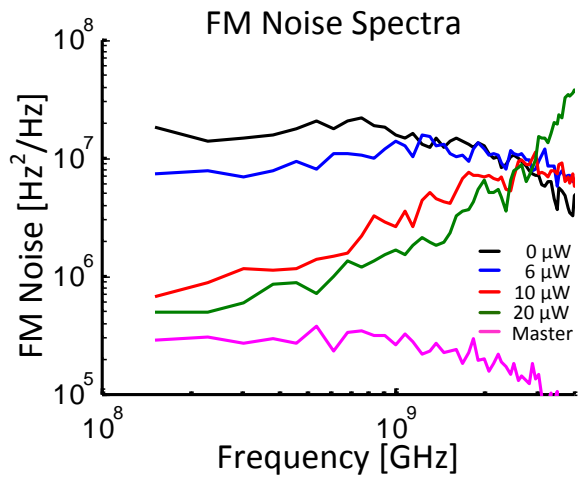

Fig. 5 Calculated FM-noise spectra of filtered comb lines for increasing injection power of the master laser. The upper and lower bounds of the FM noise are due to the gain-switched free running comb and the master laser respectively. As the power is increased, the phase noise of the individual comb lines tends to that of the master laser; though at high frequencies > 700 MHz, the FM-noise curve return to the values of the free-running case which indicates that some field component remain outside the locking range.

TABLE I Parameter definitions and values used in the simulations

| Symbol | Definition | Value |
|---|---|---|
| $A$ | Nonradiative recombination rate | $1 \times 10^9$ s$^{-1}$ |
| $B$ | Radiative recombination coefficient | $1 \times 10^{-16}$ s$^{-1}$m$^{-6}$ |
| $C$ | Auger recombination coefficient | $1 \times 10^{-41}$ s$^{-1}$m$^{-9}$ |
| $a$ | Differential gain | $8 \times 10^{-13}$ |
| $N_0$ | Carrier density at transparency | $1 \times 10^{24}$ m$^{-3}$ |
| $V$ | Volume of active section | $6 \times 10^{-17}$ m$^3$ |
| $\alpha_H$ | Linewidth enhancement factor | 3 |
| $\tau_P$ | Photon lifetime | 3 ps |
| $k_c$ | Injection coupling factor | $2 \times 10^{12}$ s$^{-1}$ |
| $\beta$ | Fraction of stimulated emission into lasing mode | 0.001 |
| $A_{eff}$ | Effective modal area | $1 \times 10^{-13}$ m$^2$ |
| $e$ | Quantum of electronic charge | $1.6 \times 10^{-19}$ C |